\documentclass{elsart}

\usepackage{natbib}
\usepackage{graphicx}
\usepackage{amssymb}

\journal{New Astronomy}

\begin{document}

\begin{frontmatter}

\title{Numerical Relativistic Hydrodynamics Based on the Total Variation
       Diminishing Scheme}

\author{Eunwoo Choi\corauthref{cor}}
\address{Department of Physics and Astronomy, Georgia State University,
         P.O. Box 4106, Atlanta, GA 30302--4106, USA}
\ead{echoi@chara.gsu.edu}
\corauth[cor]{Corresponding author.}

\author{Dongsu Ryu}
\address{Department of Astronomy and Space Science,
         Chungnam National University, Daejeon 305--764, Korea}
\ead{ryu@canopus.cnu.ac.kr}

\begin{abstract}

This paper describes a multidimensional hydrodynamic code which can be
used for studies of relativistic astrophysical flows. The code solves
the special relativistic hydrodynamic equations as a hyperbolic system
of conservation laws based on the total variation diminishing (TVD)
scheme. It uses a new set of conserved quantities and employs an
analytic formula for transformation from the conserved quantities in
the reference frame to the physical quantities in the local rest frame.
Several standard tests, including relativistic shock tubes, a relativistic
wall shock, and a relativistic blast wave, are presented to
demonstrate that the code captures discontinuities correctly and sharply
in ultrarelativistic regimes. The robustness and flexibility of the code
are demonstrated through test simulations of the relativistic
Hawley-Zabusky shock and a relativistic extragalactic jet.

\end{abstract}

\begin{keyword}

hydrodynamics \sep methods: numerical \sep relativity

\end{keyword}

\end{frontmatter}

\section{Introduction}

Many high-energy astrophysical problems involve relativistic flows,
and thus understanding relativistic flows is important for correctly
interpreting astrophysical phenomena.
For instance, intrinsic beam velocities larger than $0.9c$ are typically
required to explain the apparent superluminal motions observed in
relativistic jets in microquasars in the Galaxy \citep{mir99} as well as
in extragalactic radio sources associated with active galactic nuclei
\citep{zen97}.
In some powerful extragalactic radio sources, ejections from galactic
nuclei produce true beam velocities of more than $0.98c$.
Relativistic descriptions are also inevitable in other situations of
rapid expansion such as the early stages of supernova explosions
\citep{bur00} and the production of energetic gamma-ray bursts
\citep{mes02}.
In the general fireball model of gamma-ray bursts, the internal energy
of gas is converted into the bulk kinetic energy during
expansion and this expansion leads to relativistic outflows with high
bulk Lorentz factors $\gtrsim100$.
Since such relativistic flows are highly nonlinear and intrinsically
complex, in addition to possessing large Lorentz factors, often
studying them {\it numerically} is the only possible approach.

For numerical study of non-relativistic hydrodynamics, explicit finite
difference upwind schemes have been developed and implemented successfully.
The schemes which have been used for astrophysical researches
include the Roe scheme \citep{roe81}, the
total variation diminishing (TVD) scheme \citep{har83}, the piecewise
parabolic method (PPM) scheme \citep{col84}, and the essentially
non-oscillatory (ENO) scheme \citep{har87}.
These schemes are based on exact or approximate Riemann solvers using
the characteristic decomposition of the hyperbolic system of
hydrodynamic conservation equations.
They all are able to capture sharp discontinuities robustly in the
complex flows, and to describe the physical solution accurately.

Although the upwind schemes were originally developed for
non-relativistic hydrodynamics, some have been extended to special
relativistic hydrodynamics.
For instance, \citet{dol95} adapted the ENO scheme to one-dimensional
relativistic hydrodynamics.
They fulfilled the ENO scheme using the local characteristic approach
which depends on the local linearizion of the system of conservation
equations.
\citet{mar96} adapted the PPM scheme to one-dimensional relativistic
hydrodynamics using an exact relativistic Riemann solver to
calculate numerical fluxes at cell interfaces.
\citet{don98} and \citet{alo99} constructed multidimensional
relativistic hydrodynamic codes based on the ENO scheme and the PPM
scheme, respectively.
Reviews of various numerical approaches and test problems can be found
in \citet{mar03} and \citet{wil03}.
These works showed that the advantage of the upwind schemes, high
accuracy and robustness, are carried over to relativistic hydrodynamics.

In this paper we describe a multidimensional code for special
relativistic hydrodynamics based on the total variation diminishing
(TVD) scheme \citep{har83}.
The TVD scheme is an explicit Eulerian finite difference upwind
scheme and an extension of the Roe scheme to second-order accuracy in
space and time.
The advantage of the TVD scheme is that a code based on it is simple
and fast, and yet performs well.
A non-relativistic hydrodynamic code based the TVD scheme was
built and applied to astrophysical problems such as the large scale
structure formation of the universe by one of authors \citep{ryu93}.
The special relativistic hydrodynamic code in this paper was built
by extending the non-relativistic code.
All the components of the the non-relativistic code was kept, so the
relativistic code has the structure parallel to the non-relativistic
counterpart.
It makes the relativistic code comprehensible and easily usable.
Through tests, we demonstrate that the newly developed code for
special relativistic hydrodynamics can handle interesting astrophysical
problems involving large Lorentz factors or ultrarelativistic
regimes where energy densities greatly exceed rest mass densities.

This paper is organized as follows.
In Section 2 we describe the step by step procedures for building the code
including the basic equations, characteristic decomposition, TVD scheme,
multidimensional extension, and Lorentz transformation.
Tests are presented in Section 3.
A summary and discussion follows in Section 4.

\section{Numerical Relativistic Hydrodynamics}

\subsection{Basic Equations}

The ideal relativistic hydrodynamic equations can be written as
a hyperbolic system of conservation equations
\begin{equation}
\frac{\partial D}{\partial t}+\frac{\partial}{\partial x_j}
\left(Dv_j\right) = 0,
\end{equation}
\begin{equation}
\frac{\partial M_i}{\partial t}+\frac{\partial}{\partial x_j}
\left(M_iv_j+p\delta_{ij}\right) = 0,
\end{equation}
\begin{equation}
\frac{\partial E}{\partial t}+\frac{\partial}{\partial x_j}
\left[\left(E+p\right)v_j\right] = 0,
\end{equation}
where the equation of state is given by
\begin{equation}
p = \left(\gamma-1\right)\left(e-\rho\right).
\end{equation}
Here, $D$, $M_i$, and $E$ are the mass density, momentum density, and
total energy density in the reference frame, and $\rho$, $v_j$, and $e$
are the  mass density, velocity, and internal plus mass energy density
in the local rest frame, respectively.
In general, the adiabatic index $\gamma$ is taken as $5/3$ for
mildly relativistic cases and as $4/3$ for ultrarelativistic cases
where $e\gg\rho$.
In equations (1)--(3), the indices $i$ and $j$ run over $x$, $y$, and
$z$ and the conventional Einstein summation is used.
The speed of light is set to unity ($c\equiv1$) throughout this paper.

The quantities in the reference frame are related to those in the local
rest frame via Lorentz transformation
\begin{equation}
D = \Gamma\rho,
\end{equation}
\begin{equation}
M_i = \Gamma^2\left(e+p\right)v_i,
\end{equation}
\begin{equation}
E = \Gamma^2\left(e+p\right)-p,
\end{equation}
where the Lorentz factor is given by
\begin{equation}
\Gamma = \frac{1}{\sqrt{1-v^2}}
\end{equation}
with $v^2 = v_x^2+v_y^2+v_z^2$.

In the non-relativistic limit, the quantities $D$, $M_i$, and $E$
approach their non-relativistic counterparts $\rho^N$, $\rho^N v_i^N$,
and $E^N+\rho^Nc^2$ and equations (1)--(3) reduce to the non-relativistic
hydrodynamic equations
\begin{equation}
\frac{\partial\rho^N}{\partial t}+\frac{\partial}{\partial x_j}
\left(\rho^N v_j^N\right) = 0,
\end{equation}
\begin{equation}
\frac{\partial\rho^N v_i^N}{\partial t}+\frac{\partial}{\partial x_j}
\left(\rho^N v_i^Nv_j^N+p^N\delta_{ij}\right) = 0,
\end{equation}
\begin{equation}
\frac{\partial E^N}{\partial t}+\frac{\partial}{\partial x_j}
\left[\left(E^N+p^N\right)v_j^N\right] = 0,
\end{equation}
where the pressure is given by
\begin{equation}
p^N = \left(\gamma-1\right)\left(E^N-\frac{1}{2}\rho^N {v^N}^2\right).
\end{equation}

\subsection{Characteristic Decomposition}

Equations (1)--(3) can be written as
\begin{equation}
\frac{\partial\vec{q}}{\partial t}
+\frac{\partial\vec{F}_j}{\partial x_j} = 0
\end{equation}
with the state and flux vectors
\begin{equation}
\vec{q} = \left[\matrix{D\cr M_i\cr E}\right], \qquad
\vec{F}_j = \left[\matrix{Dv_j\cr M_iv_j+p\delta_{ij}\cr
\left(E+p\right)v_j}\right],
\end{equation}
or as
\begin{equation}
\frac{\partial\vec{q}}{\partial t}
+A_j\frac{\partial\vec{q}}{\partial x_j} = 0, \qquad
A_j = \frac{\partial\vec{F}_j}{\partial\vec{q}}.
\end{equation}
Here, $A_j$ is the $5\times5$ Jacobian matrix composed with the
state and flux vectors.
The construction of the matrix $A_j$ can be simplified by introducing
a parameter vector, $\vec{u}$, as
\begin{equation}
A_j = \frac{\partial\vec{F}_j}{\partial\vec{u}}
\frac{\partial\vec{u}}{\partial\vec{q}}.
\end{equation}
We choose the parameter vector which consists of the physical
quantities in the local rest frame,
\begin{equation}
\vec{u} = \left[\matrix{\rho\cr v_i\cr e}\right].
\end{equation}

In building an upwind code to solve a hyperbolic system of conservation
equations, the eigen-structure (eigenvalues and eigenvectors) of the
Jacobian matrix is required.
Eigen-structures for relativistic hydrodynamics in multidimensions
were previously described, for instance, in \citet{don98}.
However, the state vector in this paper is different from that of
\citet{don98}, so the eigen-structure is different.
In the following, our eigen-structure of equation (16) is presented.
We first define the specific enthalpy, $h$, and the the sound
speed, $c_s$, respectively as
\begin{equation}
h = \frac{e+p}{\rho}, \qquad
c_s^2 = \frac{\gamma p}{\rho h}.
\end{equation}

Then the eigenvalues of $A_x$ for $j=x$ are
\begin{equation}
a_1 = \frac{\left(1-c_s^2\right)v_x-\sqrt{\left(1-v^2\right)c_s^2
\left[1-v^2c_s^2-\left(1-c_s^2\right)v_x^2\right]}}{1-v^2c_s^2},
\end{equation}
\begin{equation}
a_2 = v_x,
\end{equation}
\begin{equation}
a_3 = v_x,
\end{equation}
\begin{equation}
a_4 = v_x,
\end{equation}
\begin{equation}
a_5 = \frac{\left(1-c_s^2\right)v_x+\sqrt{\left(1-v^2\right)c_s^2
\left[1-v^2c_s^2-\left(1-c_s^2\right)v_x^2\right]}}{1-v^2c_s^2}.
\end{equation}
The eigenvalues $a_{1-5}$ represent the five characteristic speeds
associated with two sound wave modes ($a_{1,5}$) and three entropy modes
($a_{2-4}$).

The complete set of the corresponding right eigenvectors
($A_x \vec{R} = a \vec{R}$) is
\begin{equation}
\vec{R}_1 = \left[\frac{1-v_xa_1}{\Gamma h\left(1-v_x^2\right)},
a_1,\frac{\left(1-v_xa_1\right)v_y}{1-v_x^2},
\frac{\left(1-v_xa_1\right)v_z}{1-v_x^2},1\right]^\mathrm{T},
\end{equation}
\begin{equation}
\vec{R}_2 = \left[\frac{-\Gamma\left(2h-1\right)v_y}{h},
0,1,0,0\right]^\mathrm{T},
\end{equation}
\begin{equation}
\vec{R}_3 = \left[\frac{\Gamma^2\left(2h-1\right)
\left(v^2-v_x^2\right)+h}{\Gamma h},v_x,0,0,1\right]^\mathrm{T},
\end{equation}
\begin{equation}
\vec{R}_4 = \left[\frac{-\Gamma\left(2h-1\right)v_z}{h},
0,0,1,0\right]^\mathrm{T},
\end{equation}
\begin{equation}
\vec{R}_5 = \left[\frac{1-v_xa_5}{\Gamma h\left(1-v_x^2\right)},
a_5,\frac{\left(1-v_xa_5\right)v_y}{1-v_x^2},
\frac{\left(1-v_xa_5\right)v_z}{1-v_x^2},1\right]^\mathrm{T}.
\end{equation}

The complete set of the left eigenvectors ($\vec{L} A_x = a \vec{L}$),
which are orthonormal to the right eigenvectors, is
\begin{equation}
\vec{L}_1 = \left[\frac{-\Gamma h\left(v_x-a_5\right)}
{\left(h-1\right)\left(a_1-a_5\right)},\Delta_{12},
\frac{-\Gamma^2\left(2h-1\right)\left(v_x-a_5\right)v_y}
{\left(h-1\right)\left(a_1-a_5\right)},
\frac{-\Gamma^2\left(2h-1\right)\left(v_x-a_5\right)v_z}
{\left(h-1\right)\left(a_1-a_5\right)},\Delta_{15}\right],
\end{equation}
\begin{displaymath}
\vec{L}_2 = \left[\frac{\Gamma hv_y}{h-1},
\frac{\left[\Gamma^2\left(2h-1\right)\left(v^2-v_x^2\right)
+h\right]v_xv_y}{\left(h-1\right)\left(1-v_x^2\right)},
\frac{\Gamma^2\left(2h-1\right)v_y^2}{h-1}+1,
\frac{\Gamma^2\left(2h-1\right)v_yv_z}{h-1},\right.
\end{displaymath}
\begin{equation}
\left.\frac{-\left[\Gamma^2\left(2h-1\right)\left(v^2-v_x^2\right)
+h\right]v_y}{\left(h-1\right)\left(1-v_x^2\right)}\right],
\end{equation}
\begin{displaymath}
\vec{L}_3 = \left[\frac{\Gamma h}{h-1},
\frac{\left[\Gamma^2\left(2h-1\right)\left(v^2-v_x^2\right)
+1\right]v_x}{\left(h-1\right)\left(1-v_x^2\right)},
\frac{\Gamma^2\left(2h-1\right)v_y}{h-1},
\frac{\Gamma^2\left(2h-1\right)v_z}{h-1},\right.
\end{displaymath}
\begin{equation}
\left.\frac{-\Gamma^2\left(2h-1\right)\left(v^2-v_x^2\right)-1}
{\left(h-1\right)\left(1-v_x^2\right)}\right],
\end{equation}
\begin{displaymath}
\vec{L}_4 = \left[\frac{\Gamma hv_z}{h-1},
\frac{\left[\Gamma^2\left(2h-1\right)\left(v^2-v_x^2\right)
+h\right]v_xv_z}{\left(h-1\right)\left(1-v_x^2\right)},
\frac{\Gamma^2\left(2h-1\right)v_yv_z}{h-1},
\frac{\Gamma^2\left(2h-1\right)v_z^2}{h-1}+1,\right.
\end{displaymath}
\begin{equation}
\left.\frac{-\left[\Gamma^2\left(2h-1\right)\left(v^2-v_x^2\right)
+h\right]v_z}{\left(h-1\right)\left(1-v_x^2\right)}\right],
\end{equation}
\begin{equation}
\vec{L}_5 = \left[\frac{-\Gamma h\left(v_x-a_1\right)}
{\left(h-1\right)\left(a_5-a_1\right)},\Delta_{52},
\frac{-\Gamma^2\left(2h-1\right)\left(v_x-a_1\right)v_y}
{\left(h-1\right)\left(a_5-a_1\right)},
\frac{-\Gamma^2\left(2h-1\right)\left(v_x-a_1\right)v_z}
{\left(h-1\right)\left(a_5-a_1\right)},\Delta_{55}\right],
\end{equation}
where the auxiliary variables are defined as
\begin{equation}
\Delta_{12} = \frac{-\left[\Gamma^2\left(2h-1\right)
\left(v^2-v_x^2\right)+1\right]\left(v_x-a_5\right)v_x}
{\left(h-1\right)\left(1-v_x^2\right)\left(a_1-a_5\right)}
+\frac{1}{a_1-a_5},
\end{equation}
\begin{equation}
\Delta_{15} = \frac{\left[\Gamma^2\left(2h-1\right)
\left(v^2-v_x^2\right)+1\right]\left(v_x-a_5\right)}
{\left(h-1\right)\left(1-v_x^2\right)\left(a_1-a_5\right)}
-\frac{a_5}{a_1-a_5},
\end{equation}
\begin{equation}
\Delta_{52} = \frac{-\left[\Gamma^2\left(2h-1\right)
\left(v^2-v_x^2\right)+1\right]\left(v_x-a_1\right)v_x}
{\left(h-1\right)\left(1-v_x^2\right)\left(a_5-a_1\right)}
+\frac{1}{a_5-a_1},
\end{equation}
\begin{equation}
\Delta_{55} = \frac{\left[\Gamma^2\left(2h-1\right)
\left(v^2-v_x^2\right)+1\right]\left(v_x-a_1\right)}
{\left(h-1\right)\left(1-v_x^2\right)\left(a_5-a_1\right)}
-\frac{a_1}{a_5-a_1}.
\end{equation}

The eigenvalues and eigenvectors of $A_y$ and $A_z$ can be obtained
by properly redefining indices.
We note that the eigenvalues are same regardless of the choice of
state or parameter vectors. But the right and left eigenvectors
are different or can be presented in different forms.

\subsection{One-Dimensional Functioning Code Based on the TVD Scheme}

The TVD scheme we employ to build one-dimensional functioning code is
practically identical to that in \citet{har83} and \citet{ryu93}.
But for completeness, the procedure is shown here.
The state vector $\vec{q}_i^n$ at the cell center $i$ at the time step
$n$ is updated by calculating the modified flux vector
$\bar{\vec{f}}_{x,i\pm1/2}$ along the $x$-direction at the cell interface
$i\pm1/2$ as follows:
\begin{equation}
L_x\vec{q}_i^n = \vec{q}_i^n-\frac{\Delta t^n}{\Delta x}
\left(\bar{\vec{f}}_{x,i+1/2}-\bar{\vec{f}}_{x,i-1/2}\right),
\end{equation}
\begin{equation}
\bar{\vec{f}}_{x,i+1/2} = \frac{1}{2}\left[\vec{F}_x(\vec{q}_i^n)
+\vec{F}_x(\vec{q}_{i+1}^n)\right]-\frac{\Delta x}{2\Delta t^n}
\sum_{k=1}^5\beta_{k,i+1/2}\vec{R}_{k,i+1/2}^n,
\end{equation}
\begin{equation}
\beta_{k,i+1/2} = Q_k(\frac{\Delta t^n}{\Delta x}a_{k,i+1/2}^n
+\gamma_{k,i+1/2})\alpha_{k,i+1/2}-\left(g_{k,i}+g_{k,i+1}\right),
\end{equation}
\begin{equation}
\gamma_{k,i+1/2} = \left\{
\begin{array}{lcl}
\left(g_{k,i+1}-g_{k,i}\right)/\alpha_{k,i+1/2} & \mathrm{for} &
\alpha_{k,i+1/2}\neq0, \\
0 & \mathrm{for} & \alpha_{k,i+1/2}=0,
\end{array}
\right.
\end{equation}
\begin{equation}
g_{k,i} = \mathrm{sign}(\tilde{g}_{k,i+1/2})\mathrm{max}
\{0,\mathrm{min}[|\tilde{g}_{k,i+1/2}|,
\mathrm{sign}(\tilde{g}_{k,i+1/2})\tilde{g}_{k,i-1/2}]\},
\end{equation}
\begin{equation}
\tilde{g}_{k,i+1/2} = \frac{1}{2}\left[Q_k(\frac{\Delta t^n}{\Delta x}
a_{k,i+1/2}^n)-\left(\frac{\Delta t^n}{\Delta x}a_{k,i+1/2}^n\right)^2
\right]\alpha_{k,i+1/2},
\end{equation}
\begin{equation}
\alpha_{k,i+1/2} = \vec{L}_{k,i+1/2}^n\cdot\left(\vec{q}_{i+1}^n
-\vec{q}_i^n\right),
\end{equation}
\begin{equation}
Q_k(x) = \left\{
\begin{array}{lcl}
x^2/4\varepsilon_k+\varepsilon_k & \mathrm{for} & |x|<2\varepsilon_k, \\
|x| & \mathrm{for} & |x|\geq2\varepsilon_k.
\end{array}
\right.
\end{equation}
Here, $k=1$ to 5 stand for the five characteristic modes.
The internal parameters $\varepsilon_k$'s are associated with numerical
viscosity, and defined for $0\leq\varepsilon_k\leq0.5$;
$\varepsilon_{1,5} = 0.1-0.3$ for the sound wave modes and
$\varepsilon_{2-4} = 0-0.1$ for the entropy modes are reasonable choices.

We note that the flux limiter in equation (42) is the min-mod limiter.
The min-mod limiter is known to be very stable but has the cost of
additional diffusion.
To reproduce sharper structures with less diffusion, other flux limiters,
such as the monotonized central difference limiter (MC limiter)
\begin{displaymath}
g_{k,i} = \mathrm{sign}(\tilde{g}_{k,i+1/2})\mathrm{max}\{0,\mathrm{min}
[\frac{1}{2}(|\tilde{g}_{k,i+1/2}|+\mathrm{sign}(\tilde{g}_{k,i+1/2})
\tilde{g}_{k,i-1/2}),2|\tilde{g}_{k,i+1/2}|,
\end{displaymath}
\begin{equation}
2\mathrm{sign}(\tilde{g}_{k,i+1/2})\tilde{g}_{k,i-1/2}]\},
\end{equation}
or the superbee limiter
\begin{displaymath}
g_{k,i} = \mathrm{sign}(\tilde{g}_{k,i+1/2})\mathrm{max}\{0,\mathrm{min}
[|\tilde{g}_{k,i+1/2}|,2\mathrm{sign}(\tilde{g}_{k,i+1/2})
\tilde{g}_{k,i-1/2}],\mathrm{min}[2|\tilde{g}_{k,i+1/2}|,
\end{displaymath}
\begin{equation}
\mathrm{sign}(\tilde{g}_{k,i+1/2})\tilde{g}_{k,i-1/2}]\},
\end{equation}
may be used; however, these limiters are more susceptible to oscillations
at discontinuities.
In the tests described in \S 3, the min-mod limiter was used.

In order to define the physical quantities at the cell interfaces, the
TVD scheme originally used the Roe's linearizion technique \citep{har83}.
Although it is possible to implement this linearizion technique in
the relativistic domain in a computationally feasible way
\citep[see][]{eul95}, there is unlikely to be a significant advantage
from the computational point of view.
Instead, we simply calculate the algebraic averages of quantities at two
adjacent cell centers to define the physical quantities at the cell
interfaces;
\begin{equation}
v_{x,i+1/2} = \frac{v_{x,i}+v_{x,i+1}}{2},~
v_{y,i+1/2} = \frac{v_{y,i}+v_{y,i+1}}{2},~
v_{z,i+1/2} = \frac{v_{z,i}+v_{z,i+1}}{2},
\end{equation}
\begin{equation}
h_{i+1/2} = \frac{h_i+h_{i+1}}{2},
\end{equation}
\begin{equation}
c_{s,i+1/2} = \left[\frac{\left(\gamma-1\right)\left(h_{i+1/2}-1\right)}
{h_{i+1/2}}\right]^{1/2}.
\end{equation}

\subsection{Multidimensional Extension}

To extend the one-dimensional code to multidimensions, the procedure
described in the previous subsection is applied separately to the $y$
and $z$-directions.
Multiple spatial dimensions are treated through the Strang-type
dimensional splitting \citep{str68}.
Then, the state vector is updated by
\begin{equation}
\vec{q}^{n+1} = L_zL_yL_x\vec{q}^n.
\end{equation}
In order to maintain second-order accuracy in time, the order of the
dimensional splitting is permuted as follows
\begin{equation}
L_zL_yL_x,~L_xL_yL_z,~L_xL_zL_y,~L_yL_zL_x,~L_yL_xL_z,~L_zL_xL_y.
\end{equation}

The time step $\Delta t^n$ is restricted by the usual Courant stability
condition
\begin{equation}
\Delta t^n = \min\left[
\frac{C_\mathrm{Cour}\Delta x}{\mathrm{max}(a_{k,i+1/2}^n)_x},
\frac{C_\mathrm{Cour}\Delta y}{\mathrm{max}(a_{k,i+1/2}^n)_y},
\frac{C_\mathrm{Cour}\Delta z}{\mathrm{max}(a_{k,i+1/2}^n)_z}\right].
\end{equation}
The Courant constant should be $C_\mathrm{Cour} < 1$.
We typically use $C_\mathrm{Cour} \lesssim 0.9$.
The time step is calculated at the beginning of a permutation sequence
and used through the complete sequence.

\subsection{Lorentz Transformation}

In the code, the conserved quantities $D$, $M_i$, and $E$ in the
reference frame are evolved in time, but the physical quantities
$\rho$, $v_j$, and $e$ in the local rest frame are needed for
fluxes to be estimated.
The quantities $\rho$, $v_j$, and $e$ can be obtained through
Lorentz transformation of equations (5)--(7) at each time step.
\citet{sch93} showed that the transformation is reduced to a single
quartic equation for $v$
\begin{equation}
f(v) = \left[\gamma v\left(E-Mv\right)-M\left(1-v^2\right)\right]^2
-\left(1-v^2\right)v^2\left(\gamma-1\right)^2D^2 = 0,
\end{equation}
where $M^2 = M_x^2+M_y^2+M_z^2$.
They also showed that the physically meaningful solution for $v$
is between the lower limit, $v_1$, and the upper limit, $v_2$,
\begin{equation}
v_1 = \frac{\gamma E-\sqrt{\left(\gamma E\right)^2
-4\left(\gamma-1\right)M^2}}{2\left(\gamma-1\right)M},~
v_2 = \frac{M}{E},
\end{equation}
and that the solution is unique.
Once $v$ is known, the quantities $\rho$, $v_j$, and $e$ can be
straightforwardly calculated from the following relations
\begin{equation}
\rho = \frac{D}{\Gamma},
\end{equation}
\begin{equation}
v_x = \frac{M_x}{M}v,~v_y = \frac{M_y}{M}v,~v_z = \frac{M_z}{M}v,
\end{equation}
\begin{equation}
e = E-M_xv_x-M_yv_y-M_zv_z.
\end{equation}

Equation (54) could be solved using a numerical procedure such as the
Newton-Raphson root-finding method, as suggested in \citet{sch93}.
A problem with the numerical approach is, however, that iterations
can fail to converge.
For instance, convergence can fail if one of the relativistic
conditions is violated due to numerical errors, e.g., $M > E$, in a cell.
This occurs mostly in extreme regimes.
In addition, we found that convergence is often slow or sometimes
fails in the limit $M \ll E$.
On the other hand, quartic equations have analytic solutions.
The general form of roots can be found in standard books such as
\citet{abr72} or on webs such as 
``http://mathworld.wolfram.com/QuarticEquation.html''.
Although it is too complicated to prove analytically, we found
numerically that for the physical meaningful values of $v$ and $c_s$,
$v < 1$ and $c_s < \sqrt{\gamma-1}$, among the four roots of equation
(54), two are complex and the other two are real.
While the smaller real root is smaller than the lower limit $v_1$,
the larger real root is between the two limits $v_1$ and $v_2$.
So the larger real root is the one we are looking for, and we use
its analytic formula in our code.
The advantages of the analytic approach are obvious.
It always gives a solution we are looking for,
and it is easier to predict and deal with unphysical situations if
one of the relativistic conditions is violated due to numerical errors.

\section{Numerical Tests}

\subsection{Relativistic Shock Tube}

\begin{figure}
\includegraphics[scale=0.7]{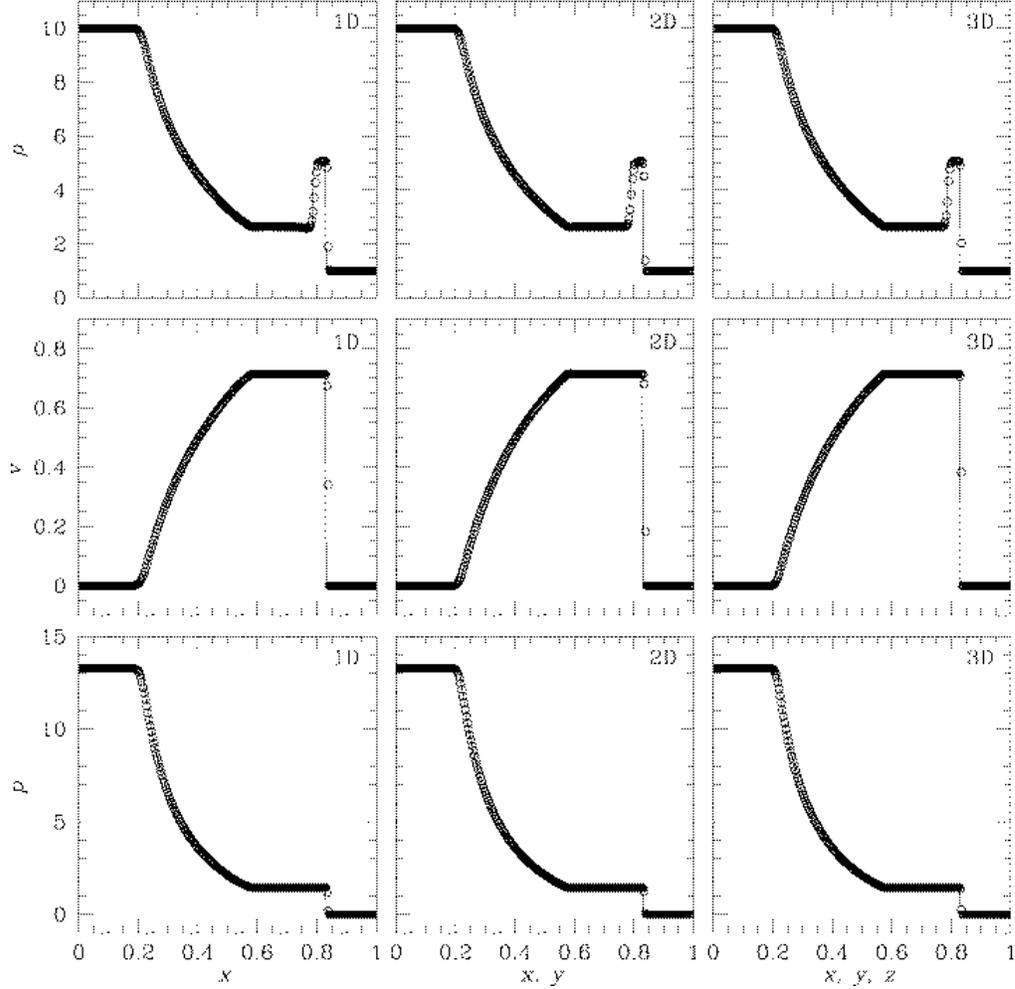}
\caption{(a) 1D, 2D, and 3D mildly relativistic shock tube tests.
The calculations have been done with the initial states in equation (59)
using $256$, $256^2$, and $256^3$ cells.
The numerical solutions (open circles) and the exact solutions (solid
lines) are plotted at $t = 0.4$, $0.4\sqrt2$, and $0.4\sqrt3$.
(b) 1D, 2D, and 3D highly relativistic shock tube tests.
The calculations have been done with the initial states in equation (60)
using $256$, $256^2$, and $256^3$ cells.
The numerical solutions (open circles) and the exact solutions (solid
lines) are plotted at $t = 0.4$, $0.4\sqrt2$, and $0.4\sqrt3$.}
\label{fig1}
\end{figure}

We have performed two sets of relativistic shock tube tests in the one,
two, and three-dimensional computational boxes with $x = [0,1]$,
$y = [0,1]$, and $z = [0,1]$.
Initially two different physical states are set up perpendicular to the
direction along which waves propagate; along the $x$-axis in the
one-dimensional calculation, along the diagonal line connecting $(0,0)$
and $(1,1)$ in the two-dimensional calculation, and along the diagonal
line connecting $(0,0,0)$ and $(1,1,1)$ in the three-dimensional
calculation.
The initial states of the first test are
\begin{equation}
\left(\rho,v_x,v_y,v_z,p\right) = \left\{
\begin{array}{ll}
\left(10,0,0,0,13.3\right) & 0\leq x,~\left(x+y\right)/2,~
\left(x+y+z\right)/3\leq1/2, \\
\left(1,0,0,0,10^{-6}\right) & 1/2<x,~\left(x+y\right)/2,~
\left(x+y+z\right)/3\leq1.
\end{array}
\right.
\end{equation}
The initial states of the second test are
\begin{equation}
\left(\rho,v_x,v_y,v_z,p\right) = \left\{
\begin{array}{ll}
\left(1,0,0,0,10^3\right) & 0\leq x,~\left(x+y\right)/2,~
\left(x+y+z\right)/3\leq1/2, \\
\left(1,0,0,0,10^{-2}\right) & 1/2<x,~\left(x+y\right)/2,~
\left(x+y+z\right)/3\leq1.
\end{array}
\right.
\end{equation}
In equations (59) and (60), the expressions within inequalities are
for one, two, and three dimensions, respectively.
The first test involves a mildly relativistic flow and the second test
involves a highly relativistic flow.
In both tests, we assume the adiabatic index $\gamma = 5/3$ and the
outflow condition is used for the $x$, $y$, and $z$-boundaries.
Both tests were previously considered by several authors
\citep[e.g.,][]{mar96}.
The estimation of accuracy was done by comparing the numerical solutions
with the exact solutions described in \citet{tho86} and \citet{mar94}.
In Figures \ref{fig1}(a) and (b), our numerical solutions are shown as
open circles and the exact solutions are represented by solid lines.

Figure \ref{fig1}(a) shows the mildly relativistic shock tube test done
using $256$, $256^2$, and $256^3$ cells with a Courant constant
$C_\mathrm{Cour} = 0.9$ and the parameters $\varepsilon_{1,5} = 0.1$ and
$\varepsilon_{2-4} = 0$.
The plots of one, two, and three-dimensions correspond to times
$t = 0.4$, $0.4\sqrt2$, and $0.4\sqrt3$, respectively.
Structures such as the shock front, contact discontinuity and
rarefaction wave are accurately produced.
There are actually slight improvements in accuracy in the
multidimensional calculations.
Figure \ref{fig1}(b) shows the highly relativistic shock tube test done
again using $256$, $256^2$, and $256^3$ cells with a Courant constant
$C_\mathrm{Cour} = 0.6$ and the parameters $\varepsilon_{1,5} = 0.1$ and
$\varepsilon_{2-4} = 0$.
The plots of one, two, and three-dimensions correspond to times
$t = 0.4$, $0.4\sqrt2$, and $0.4\sqrt3$, respectively.
The flow is more extreme, but the structure is correctly reproduced
without spurious oscillations.
But in the rest mass density profile the peak does not reach the value
of the exact solution due to the coarseness of computational cells.
According to our tests, in a one-dimensional calculation, the peak can
be accurately reproduced when $2048$ numerical cells are used.
There are also improvements in accuracy in the multidimensional
calculations.

For a more quantitative comparison, we have calculated the norm errors of
the rest mass density, velocity, and pressure for different dimensions.
The errors shown in Table \ref{tab1} are calculated at the same times as
in Figure \ref{fig1}.
The errors are gradually reduced as the dimensionality increases and
demonstrate a good agreement between the numerical and exact solutions.
Note that the values of $\|E(p)\|$ exceeding unity are still acceptable
because these are from the initial large value of pressure.

\subsection{Relativistic Wall Shock}

A one-dimensional relativistic wall shock test has been performed in
the computational box of $x = [0,1]$.
Initially a gas with extreme velocity occupying all numerical cells
propagates along the $x$-axis against a reflecting wall placed at
$x = 1$.
As the gas hits the wall, it is compressed and heated and eventually a
reverse shock is generated.
The initial condition of this test is
\begin{equation}
\left(\rho,v_x,v_y,v_z,p\right) = \left(1,0.999999,0,0,10^{-4}\right) \qquad
0\leq x\leq1.
\end{equation}
The adiabatic index $\gamma = 5/3$ is assumed and the inflow boundary
condition is used at $x = 0$.
It is another test which was widely used by several authors
\citep[e.g.,][]{don98}.

The relativistic jump condition for strong shocks with negligible preshock
pressure is given by \citet{bla76}
\begin{equation}
v_s = -\frac{\left(\gamma-1\right)\Gamma v}{\Gamma+1},
\end{equation}
\begin{equation}
\rho^* = \rho\frac{\gamma\Gamma+1}{\gamma-1},
\end{equation}
\begin{equation}
v^* = 0,
\end{equation}
\begin{equation}
p^* = \rho\left(\Gamma-1\right)\left(\gamma\Gamma+1\right).
\end{equation}
Here, $v_s$ is the shock velocity and the superscript $\ast$ represents
the postshock quantities, while the quantities without any superscript
refer to the preshock gas.

Figure \ref{fig2} shows the structure at $t = 0.75$ when the reverse shock
is located at $x = 0.5$.
The calculation has been done using $512$ computational cells with a
Courant constant $C_\mathrm{Cour} = 0.9$ and the parameters
$\varepsilon_{1,5} = 0.3$ and $\varepsilon_{2-4} = 0.1$.
The numerical solution is drawn with open circles and the exact solution
is represented by solid lines.
The numerical and exact solutions match exactly without any oscillation
or overshoot in the rest mass density, velocity, and pressure profiles.

With different inflow velocities, we have calculated the mean errors
in the rest mass density, velocity, and pressure.
The errors are calculated for the same time as in Figure \ref{fig2} and
given in Table \ref{tab2}.
Note that the order of the mean errors is $10^{-3}$, and that the
accuracy does not depend systematically on the investigated Lorentz
factor.
The mean error in the rest mass density is $\lesssim0.5\%$ for all the
Lorentz factors and about $0.25\%$ for the maximum Lorentz factor.
This accuracy is comparable to or better than that of other published
upwind scheme codes.

\subsection{Relativistic Blast Wave}

The propagation of a relativistic blast wave has been tested in the
two-dimensional computational box with $x = [0,1]$ and $y = [0,1]$.
A gas of high density and pressure is initially confined in a spherical
region and the subsequent explosion is allowed to evolve.
This makes a spherical blast wave propagate outward.
The initial condition of this test is
\begin{equation}
\left(\rho,v_x,v_y,v_z,p\right) = \left\{
\begin{array}{ll}
\left(10,0,0,0,10^3\right) & 0\leq\sqrt{x^2+y^2}\leq1/2, \\
\left(1,0,0,0,1\right) & \mathrm{outside}.
\end{array}
\right.
\end{equation}
The adiabatic index is taken to be $\gamma = 4/3$ and the reflecting and
outflow boundary conditions are used.

The calculation has been done using $512^2$ cells with a Courant
constant $C_\mathrm{Cour} = 0.6$ and the parameters
$\varepsilon_{1,5} = 0.1$ and $\varepsilon_{2-4} = 0$.
To test the symmetry properties of the code, the calculation has been
stopped before a reverse shock reaches the inner reflecting boundary.
Figure \ref{fig3} shows the profiles of the rest mass density, velocity, and
pressure measured along the diagonal line connecting $(0,0)$ and $(1,1)$
at $t = 0.7$.
The spherical blast wave successfully propagates to a larger radius, and we
have found that all structures in it preserve the initial symmetry.

\subsection{Relativistic Hawley-Zabusky Shock}

In order to test the applicability of the code to complex relativistic
flows, we have performed a two-dimensional test simulation of the
relativistic version of the Hawley-Zabusky shock.
The test was originally suggested by \citet{haw89} for non-relativistic
hydrodynamics.
Almost the same physical values as in the original paper are used
here.
Initially a plane-parallel shock with a Mach number $1.2$ propagates
along the $x$-axis into two regions of different density.
The regions are separated by oblique discontinuity whose inclination is
$30^\circ$ with respect to the $x$-axis.
The density jumps three times across the discontinuity.
The initial configuration is summarized as
\begin{equation}
\left(\rho,v_x,v_y,v_z,p\right) = \left\{
\begin{array}{ll}
\left(1,0.6,0,0,0.48\right) & 0\leq x\leq1/16,~0\leq y\leq1, \\
\left(1,0,0,0,0.48\right) & 1/16<x\leq\sqrt3y+1/4,~0\leq y\leq1, \\
\left(3,0,0,0,0.48\right) & \mathrm{outside}.
\end{array}
\right.
\end{equation}
The adiabatic index $\gamma = 1.4$ is used.
The inflow and outflow conditions are used at the $x$-boundaries and
the reflecting condition is used at the $y$-boundaries.

The simulation has been done in the two-dimensional computational box
with $x = [0,8]$ and $y = [0,1]$ using a uniform numerical grid of
$2048\times256$ cells.
A Courant constant $C_\mathrm{Cour} = 0.9$ and the parameters
$\varepsilon_{1,5} = 0.1$ and $\varepsilon_{2-4} = 0$ were used.
We have simulated this test until $t = 20$ in order to see the long
term evolution.
The passage of the planar shock through the discontinuity causes the
Kelvin-Helmholtz instability to occur along the discontinuity
and end up with formation of vortices.
The vortices roll up, interact, and merge during the simulation; the
detailed morphology and the number of vortices formed are somewhat
sensitive to numerical resolution.
Figure \ref{fig4} shows the gray-scale images of the rest mass density at
different times ($t = 2$, $11$, and $20$).
Because all the structures are dragged to the right boundary as time
goes on, only the left, middle, and right half of the computational
box are shown at $t = 2$, $11$, and $20$, respectively.
The vortices along the discontinuity are clearly formed and overall
the morphology is similar to that of the non-relativistic simulation.

\subsection{Relativistic Extragalactic Jets}

Finally, in order to test the applicability of the code to realistic
relativistic flows, we have simulated a two-dimensional relativistic
extragalactic jet propagating into homogeneous medium.
The relativistic jet inflows with a velocity $0.99$ to the
computational box of $x = [0,4]$ and $y = [0,1]$.
The jet has initially radius $1/8$ ($32$ cells) and Mach number $8.76$.
The density ratio of the jet to the ambient medium is $0.1$ and the
pressure of the jet is in equilibrium with that of the ambient medium.
The initial condition for jet inflow and ambient medium is summarized as
\begin{equation}
\left(\rho,v_x,v_y,v_z,p\right) = \left\{
\begin{array}{ll}
\left(1,0.99,0,0,0.1\right) & 0\leq x\leq1/32,~0\leq y\leq1/8, \\
\left(10,0,0,0,0.1\right) & \mathrm{outside}.
\end{array}
\right.
\end{equation}
The adiabatic index $\gamma = 4/3$ is used.
The inflow and outflow conditions are used at the $x$-boundaries and
the reflecting and outflow conditions are used at the $y$-boundaries.

The simulation has been done using a uniform numerical grid of
$1024\times256$ cells with a Courant constant $C_\mathrm{Cour} = 0.3$
and the parameters $\varepsilon_{1,5} = 0.3$ and
$\varepsilon_{2-4} = 0.1$.
Figure \ref{fig5} shows the gray-scale images of logarithm of the rest mass
density, pressure, and Lorentz factor at $t = 5$ when the bow shock
reaches the right boundary.
We can clearly see the dominant structures of bow shock, working
surface, contact discontinuity, and cocoon.
It is clear that the internal structure of the relativistic jet is less
complex compared to that of a non-relativistic jet due to the effects of
high Lorentz factor.
The overall morphology and dynamics of our simulation match roughly
with those of previous works, e.g., \citet{dun94}, although the initial
conditions and the plotted epoch are different.

\section{Summary and Discussion}

A multidimensional code for special relativistic hydrodynamics was
described. It differs from previous codes in the following aspects:
1) It is based on the total variation diminishing (TVD) scheme
\citep{har83}, which is an explicit Eulerian finite difference upwind
scheme and an extension of the Roe scheme to second-order accuracy in
space and time.
2) It employs a new set of conserved quantities, and so the paper
describes a new eigen-structure for special relativistic hydrodynamics.
3) For the Lorentz transformation from the conserved quantities in the
reference frame to the physical quantities in the local rest frame,
an analytic formula is used.

To demonstrate the performance of the code, several tests were presented,
including relativistic shock tubes, a relativistic wall shock, a
relativistic blast wave, the relativistic version of the Hawley-Zabusky
shock, and a relativistic extragalactic jet.
The relativistic shock tube tests showed that the code clearly resolves
mildly relativistic and highly relativistic shocks within $2-4$
numerical cells, although it requires more cells for resolving contact
discontinuities.
The relativistic wall shock test showed that the code correctly
captures very strong shocks with very high Lorentz factors.
The relativistic blast wave test showed that blast waves propagate
through ambient medium while preserving the symmetry.
The test simulations of the relativistic version of the Hawley-Zabusky
shock and a relativistic extragalactic jet proved the robustness and
flexibility of the code, and that the code can be applied to studies of
practical astrophysical problems.

The strong points of the new code include the following:
1) Based on the TVD scheme, the code is simple and fast.
The core routine of the TVD relativistic hydrodynamics is only about
300 lines long in the three-dimensional version.
It runs only about $1.5 - 2$ times slower than the non-relativistic
counterpart (per time step).
Yet, tests have shown that the code is accurate and reliable enough to
be suited for astrophysical applications.
In addition, the use of an analytic formula for Lorentz transformation
makes the code robust, so it ran for all the tests we have performed
without failing to converge.
2) The code has been built in a way to be completely parallel to the
non-relativistic counterpart. So it can be easily understood and used,
once one is familiar with the non-relativistic code.
In addition, the techniques developed for the non-relativistic code
such as parallelization can be imported transparently.

Finally, the code is currently being applied for studies of
relativistic jet interactions with inhomogeneous external media and
turbulence of relativistic flows. The results will be reported in
separate papers.

\ack

EC is grateful to Paul Wiita for his advice and comments on this work.
EC was supported by the GSU College of Arts and Sciences, and
by Research Program Enhancement funds to the Program in Extragalactic
Astronomy.
DR was supported by the KOSEF grant R01-2004-000-10005-0.

\clearpage

\begin{table}
\caption{Norm errors for the relativistic shock tube tests}
\label{tab1}
\begin{tabular}{rlccc}
\hline\hline
 RST & $n_\mathrm{cell}$ & $\|E(\rho)\|$ & $\|E(v)\|$ & $\|E(p)\|$ \\
\hline
 (a) 1D & $256$   & 1.1688E$-$01 & 6.0952E$-$02 & 9.3517E$-$02 \\
     2D & $256^2$ & 1.1264E$-$01 & 6.0586E$-$02 & 9.6789E$-$02 \\
     3D & $256^3$ & 9.1309E$-$02 & 5.8222E$-$02 & 8.7047E$-$02 \\
 (b) 1D & $256$   & 1.7506E$-$01 & 2.6591E$-$02 & 5.2191E$+$00 \\
     2D & $256^2$ & 1.6375E$-$01 & 1.9552E$-$02 & 4.3126E$+$00 \\
     3D & $256^3$ & 1.3840E$-$01 & 1.3533E$-$02 & 2.8773E$+$00 \\
\hline
\end{tabular}
\end{table}

\clearpage

\begin{table}
\caption{Mean errors for the relativistic wall shock tests}
\label{tab2}
\begin{tabular}{lrccc}
\hline\hline
 $v$ & $\Gamma$ & $\bar{E}(\rho)$ & $\bar{E}(v)$ & $\bar{E}(p)$ \\
\hline
 0.9      & 2.3   & 4.7423E$-$03 & 3.1483E$-$03 & 5.8100E$-$03 \\
 0.99     & 7.1   & 3.1938E$-$03 & 2.3634E$-$03 & 2.5168E$-$03 \\
 0.999    & 22.4  & 3.1876E$-$03 & 2.6687E$-$03 & 2.5015E$-$03 \\
 0.9999   & 70.7  & 5.0532E$-$03 & 4.1790E$-$03 & 3.8529E$-$03 \\
 0.99999  & 223.6 & 2.8425E$-$03 & 2.4914E$-$03 & 2.1466E$-$03 \\
 0.999999 & 707.1 & 2.4855E$-$03 & 2.0237E$-$03 & 1.8747E$-$03 \\
\hline
\end{tabular}
\end{table}

\clearpage

\clearpage

\begin{figure}
\includegraphics[scale=0.7]{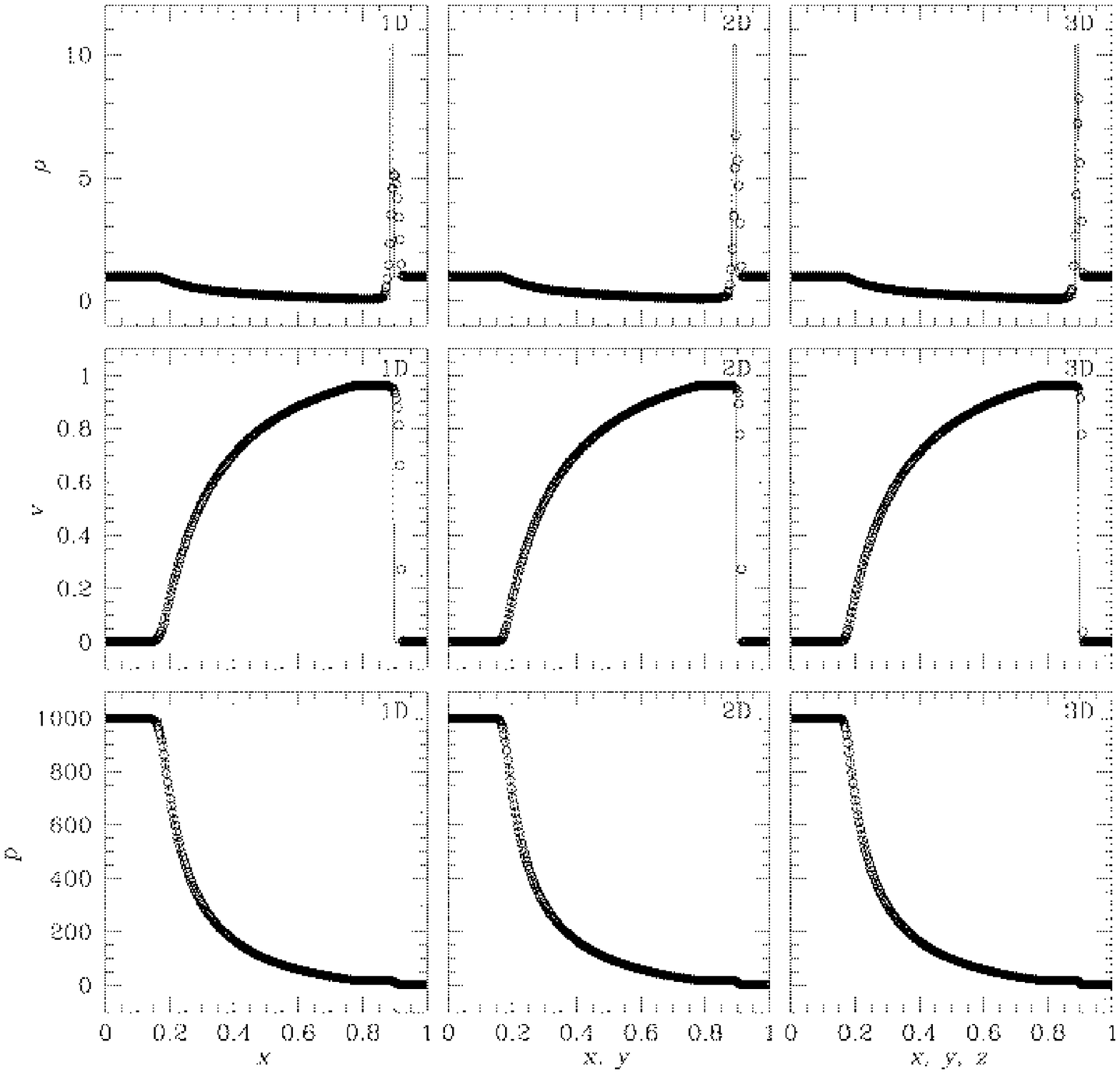}
\end{figure}

\clearpage

\begin{figure}
\includegraphics[scale=0.7]{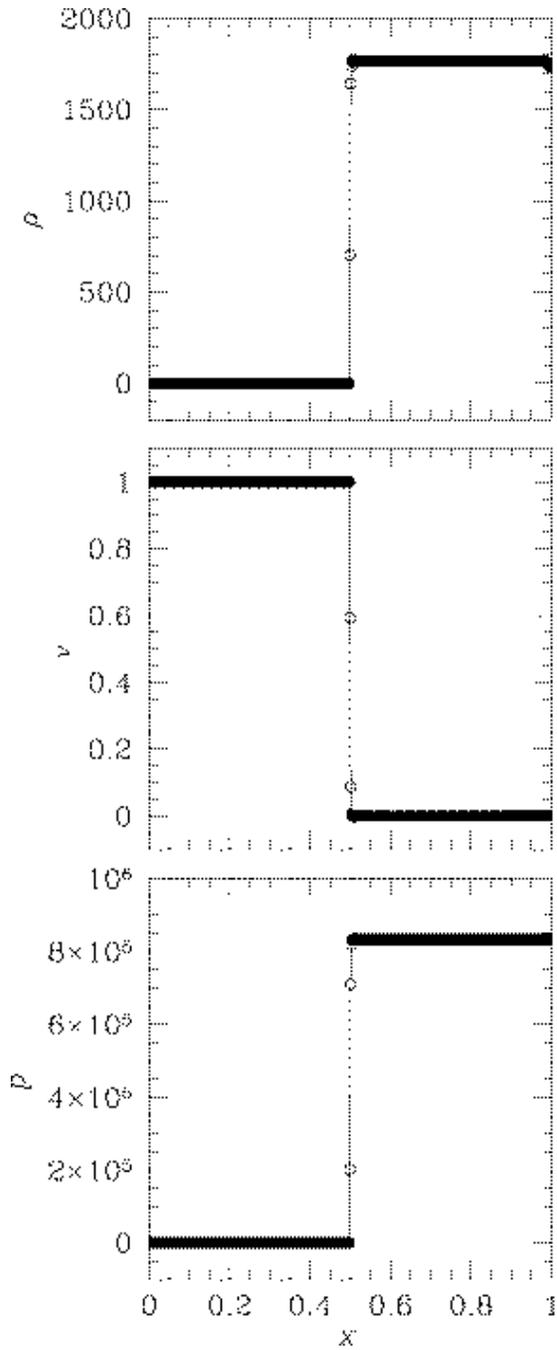}
\caption{One-dimensional relativistic wall shock test.
The calculation has been done with the initial states in equation (61)
using $512$ cells.
The numerical solution (open circles) and the exact solution (solid
lines) are plotted at $t = 0.75$.}
\label{fig2}
\end{figure}

\clearpage

\begin{figure}
\includegraphics[scale=0.7]{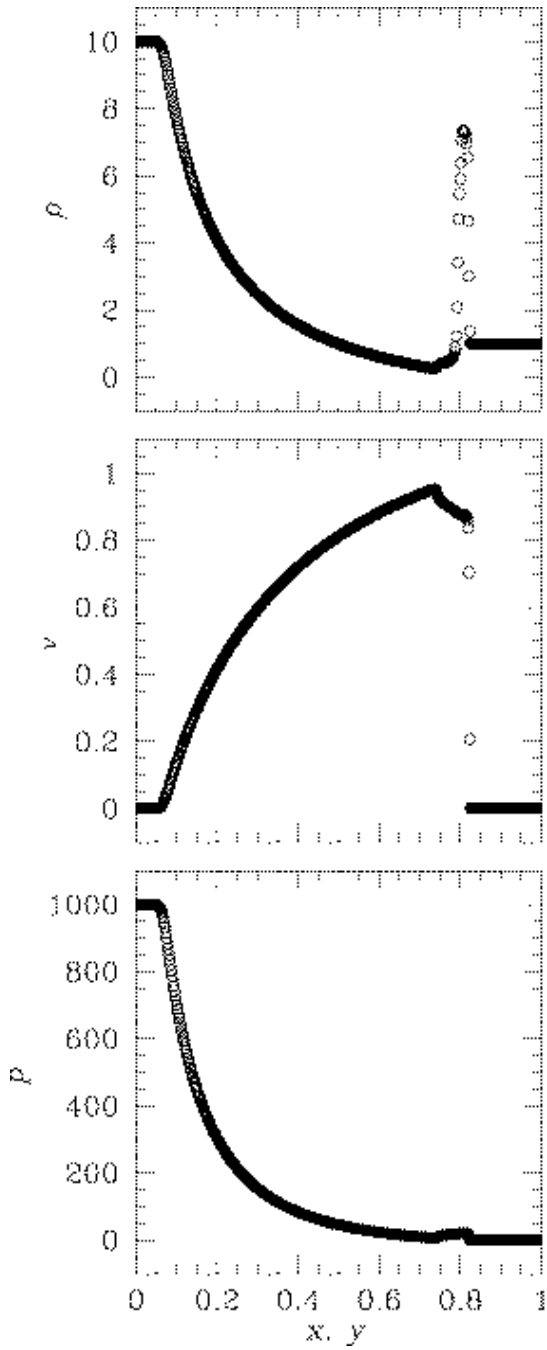}
\caption{Two-dimensional relativistic blast wave test.
The calculation has been done with the initial states in equation (66)
using $512^2$ cells.
The numerical solution (open circles) are plotted at $t = 0.7$.}
\label{fig3}
\end{figure}

\clearpage

\begin{figure}
\includegraphics[scale=0.8]{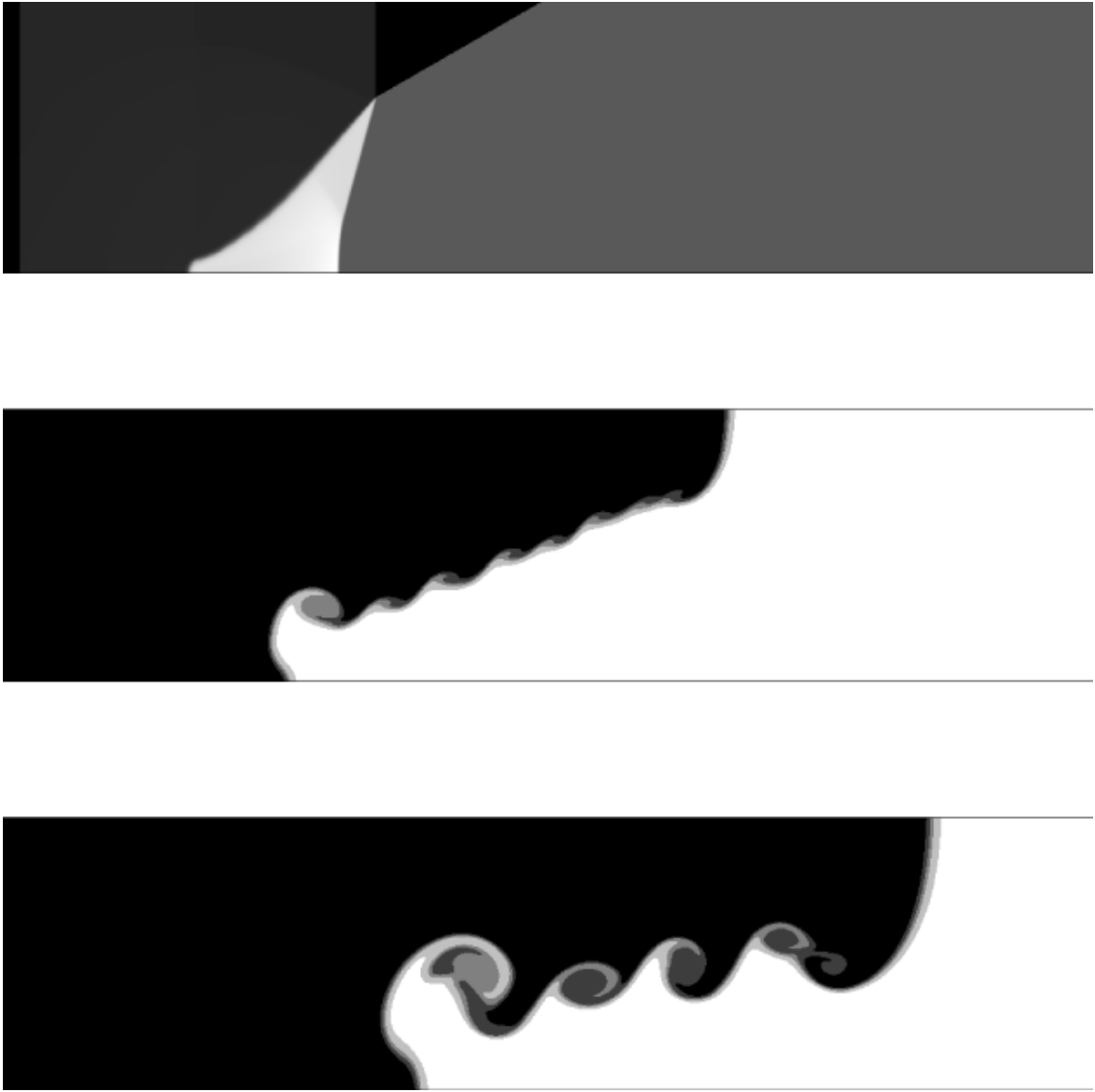}
\caption{Two-dimensional relativistic Hawley-Zabusky shock.
The simulation has been carried out with the initial configurations in
equation (67) using $2048\times256$ cells.
Gray-scale images show the rest mass density at $t = 2$, $11$, and
$20$ (top to bottom), using linear scales that range from $1.0$ (black)
to $6.75$ (white).}
\label{fig4}
\end{figure}

\clearpage

\begin{figure}
\includegraphics[scale=0.8]{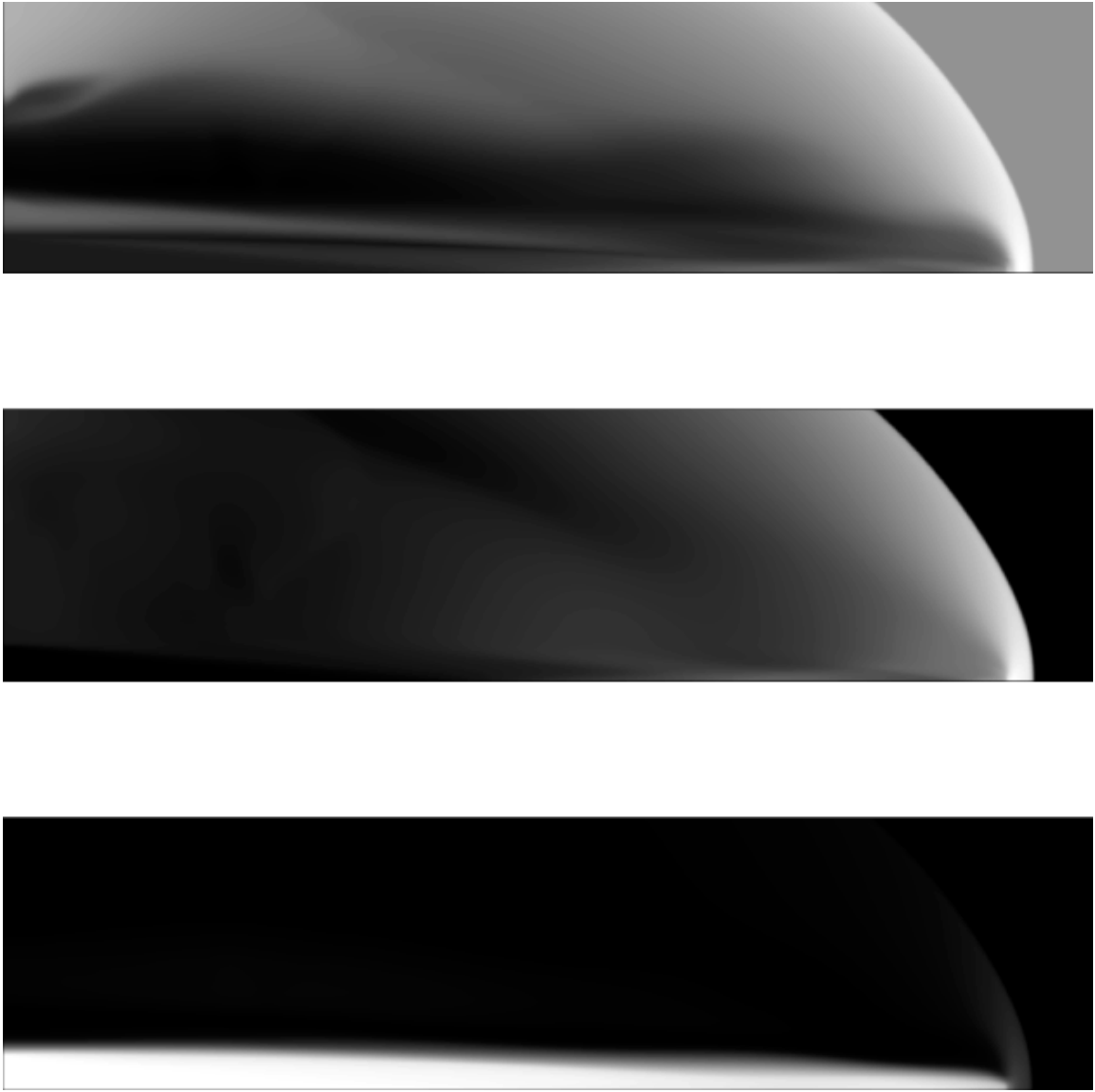}
\caption{Two-dimensional relativistic extragalactic jet.
The simulation has been carried out with the initial conditions in
equation (68) using $1024\times256$ cells.
Gray-scale images show logarithm of the rest mass density, pressure,
and Lorentz factor (top to bottom) at $t = 5$, using logarithmic scales
that range from $-0.28$ (black) to $1.98$ (white) for log(density),
$-1.0$ (black) to $1.30$ (white) for log(pressure), and $0$ (black) to
$0.85$ (white) for log(Lorentz factor).}
\label{fig5}
\end{figure}

\end{document}